# Detecting Discussions of Technical Debt

Ipek Ozkaya, Zachary Kurtz, Robert L. Nord
Software Engineering Institute
Carnegie Mellon University
Pittsburgh, PA USA
{ozkaya, rn}@sei.cmu.edu, ztkurtz@cert.org

Raghvinder S. Sangwan, Satish M. Srinivasan
School of Graduate Professional Studies
Penn State University
Malvern, PA USA
{rsangwan, sus64}@psu.edu

*Abstract*—Technical debt (TD) refers to suboptimal choices during software development that achieve short-term goals at the expense of long-term quality. Although developers often informally discuss TD, the concept has not yet crystalized into a consistently applied label when describing issues in most repositories. We apply machine learning to understand developer insights into TD when discussing tickets in an issue tracker. We generate expert labels that indicate whether discussion of TD occurs in the free text associated with each ticket in a sample of more than 1,900 tickets in the Chromium issue tracker. We then use these labels to train a classifier that estimates labels for the remaining 475,000 tickets. We conclude that discussion of TD appears in about 16% of the tracked Chromium issues. If we can effectively classify TD-related issues, we can focus on what practices could be most useful for their timely resolution.

## I. INTRODUCTION

Technical debt (TD) is a concept that reflects the implied cost of rework attributable to suboptimal solutions during software development to achieve short-term goals at the expense of the long-term quality of a software product [28]. Once incurred, TD continues to accumulate as a system evolves [5]. Previous studies find that developers discuss TD issues and agree that TD is a problem worth actively managing [7] [8] [22] [39]. Issue trackers are a natural venue for discussing TD since developers use issue trackers to coordinate task priorities. However, we have only recently started seeing developers explicitly use the phrase "technical debt" or similar terms such as "design debt" or "architectural smells." A simple search for "technical debt" on GitHub dating from 2010 returns 17,000 commits and 12,000 issues.

Several previous studies use developer discussions or code comments to automatically identify TD. Codabux and Williams [15] used the Jira issue tracking system to predict the TD-proneness of object-oriented classes in Apache Hive code using a Bayesian network. Their work limits TD items to defect- and change-prone classes extracted from Jira. Huang et al. [22] and Maldonado et al. [33] use machine learning to identify keywords that indicate TD in developer comments embedded in the source code from several open-source systems.

This paper introduces techniques that help detect TD automatically and estimate its prevalence, first steps toward systematizing discussions of TD in issue trackers. To our knowledge, our study is the first to use machine learning to detect TD in an issue tracker.

We manually labeled tickets in the Chromium issue tracker to determine whether the discussions in each ticket indicated TD. We used this labeled data to estimate the prevalence of TD in tracked Chromium issues and to train a classifier to identify unlabeled tickets that contain TD. In more detail, our contributions are as follows:

- We generated labels for 1,934 tickets from the Chromium issue tracker; 441 of these tickets show strong evidence of TD. We are making these labels publicly available to support future research [13].

- We estimated that developers identify TD as an important underlying factor in 14.5–17.1% of tracked Chromium tickets. This estimate is adjusted (making it more conservative) for selection bias in our labeled sample.

- We trained a gradient boosting machine to automatically determine whether a Chromium ticket concerns TD. On a pure holdout test set, our classifier performed significantly better than a naïve keyword search in terms of precision, recall, and area under the receiver-operator characteristic (AUROC) curve, after adjusting for sampling bias in our labeled data.

Accurately detecting discussions of TD issues manually requires analysis of multiple developer comments in the form of lengthy unstructured text. Our engine for automatically detecting TD therefore extensively uses text mining, an approach increasingly used in software engineering studies that focus on identifying and classifying bugs [3] [20]. Text mining has been found to improve both bug classification accuracy [38] and problem report classification [29].

Both new labeled data and subsequent quantitative analysis that we present in this paper raise the profile of TD as a software engineering construct of importance comparable to "security" and "resilience." Particularly, because each of these constructs is highly complex and context dependent, labeled data sets and machine learning tools are critically needed to help developers locate, discuss, and address TD throughout a software development life cycle.

A more-specific motivation for this study is to improve time to resolution and avoid unintended consequences due to misreporting of TD issues. For instance, several studies suggest that bug reports that include certain bug descriptors tend to get addressed sooner [6] [21] [23] [45]. Some issue trackers make it easier to report such descriptors by including fields or tags for them. Existing issue trackers, however, generally do not contain fields adequate for reporting of TD in a standardized way. In the present study, we report estimates of the prevalence of TD in Chromium tickets, since the prevalence of an issue is an important consideration for deciding whether to designate a field for it.

The remainder of the paper is organized as follows. Section II presents the Chromium issue tracker data by providing exemplars of tickets that contain discussions of TD and describing how we gathered the Chromium data and how we subsequently decided which tickets to label. Section III identifies ticket features of potential relevance to TD, including features derived from free text. Section IV describes our main model that uses ticket features to automatically detect tickets that concern TD. Section V evaluates the performance of our main model and estimates the overall prevalence of TD in Chromium, Section VI discusses threats to validity, Section VII describes related work, and Section VIII concludes the paper with thoughts on future directions.

## II. CHROMIUM ISSUE TRACKER DATA SET

Issue trackers serve as an entry point for communicating TD because developers use them to manage task priorities. The Chromium issue tracker [14] in particular contains rich examples of discussions about TD and how it could be resolved. The Chromium web browser project also serves as a good case study for TD for several reasons. It is a large project, with many sub-teams working independently since the project began in 2008, creating abundant opportunity for TD to arise. The Chromium developer community follows relatively robust issue-tracker practices, which means that ticket data is often sufficient to disambiguate TD from other concerns. Finally, Chromium is open source, making much of the associated data publically available.

### A. Introduction to Discussions of Technical Debt in Chromium

The term "technical debt" first appears in the Chromium issue tracker in 2010, though developers have been discussing the concept and its management throughout the lifespan of the project. Here is a sample of developer comments in that ticket:

> *[Chromium #43780] One might consider this a technical debt paydown bug. However, feel free to reprioritize....*
> *Backup sockets were committed conditionally on them being refactored to the "right" place (10/18/2010)...*
> *Looks like the statements about the code are still true. (08/17/2017)*
>
> *[https://bugs.chromium.org/p/chromium/issues/detail?id=43780]*

The discussions clearly demonstrate a short-term trade-off that was taken at the time with the understanding that it would be refactored, but this ticket remains open as of this writing.

In other examples, we see indications of a practice emerging in dealing with TD as Chromium developers discuss the tickets:

> *[Chromium #243948] Paying off technical debt becomes a higher priority, not lower, when in those rare cases it must be deferred. Tests are not a 'nice to fix' feature. Raising to Pri-1.*
>
> *[https://bugs.chromium.org/p/chromium/issues/detail?id=243948]*

It is straightforward to spot TD issues when the developers explicitly refer to them as such. Previous work has demonstrated that discussions of TD often involve more convoluted design issues and hard-to-trace changes [17] [33]. The challenge is to locate TD when it is not explicitly discussed. For instance, here is a ticket that experts classified as TD.

> *[Chromium #442327] 1) Make sure all JNI registration functions are autogenerated by the JNI generator. Currently a few are manual and therefore must be called even when native exports are in use.*
> *2) Make the JNI generator emit both manual registration functions \*and\* native exports...*
> *... Factor out the code which generates the native export stub name for a given native function, previously duplicated in two places, and also use it in a third place: when generating the table of method registrations...*
> *adding ... as blocker because this is still not quite working as intended (though it's functional)*
>
> *[https://bugs.chromium.org/p/chromium/issues/detail?id=442327]*

The discussion in this ticket indicates that the developers recognize the limitations of manual registration and the design consequences to change the system with a configuration-time flag. The TD grows as the refactoring is postponed.

Some tickets, however, are especially challenging to label, such as Chromium ticket #507796:

> *[Chromium 507796] This is just a first step to make sure the code is being exercised. It's been tested locally but only on this configuration. Some more work might be needed to get this working in non-GN builds. Further refactoring of the Telemetry dependencies will occur in follow-on CLs....Unfortunate that that build breakage wasn't caught. ... let me know if you have any trouble diagnosing what went wrong. I don't know why so many of the other isolates would complain about crashpad_database_util not having been built.*
>
> *[https://bugs.chromium.org/p/chromium/issues/detail?id=507796]*

Experts initially labeled this as not TD since it focuses on alignment of unit tests, a routine task after changes have been made. Upon further consideration, they correctly identified the design concepts by cueing on the build dependencies discussed, and they relabeled this ticket as TD.

Occasionally, related words or concepts that reflect frustration like "I don't understand," "we are getting nagged," or "workaround" suggest that the issues affecting software developers are symptomatic of TD. Other phrases that refer to the consequence of design changes such as "increasingly complex," "consequence of refactoring," or "should not have been hard coded" are suggestive of TD as well.

### B. Scraping the Issue Tickets

We used the bs4 Python module (aka "Beautiful Soup") to build a web scraper for the Chromium issue tracker and used it to gather metadata and free text associated with 477,000 tickets, a subset of the 712,000 tickets that were opened between August 30, 2008, and April 14, 2017. The vast majority of the 235,000 missing tickets in this time range were unavailable to us due to permissions controls, since some Chromium issues relate to internal business at Google. Our web scraper gathered numerous fields associated with each ticket; Section III describes these fields.

*C. Manually Labeling Tickets*

*1) Rubric:* To label each ticket, our expert raters used the rubric from Bellomo et al. [8], which defines TD as "design work related to software units that may carry present or anticipated accumulation of extra work." The rubric includes recognition of TD in discussions where developers articulate concerns related to TD even when they do not necessarily use the terms "technical" or "debt." The rubric takes the form of a decision tree that requires expert judgement at each branch. Key decision points include deciding whether the discussion contains specific evidence of a development artifact, a type of system improvement or defect, design limitations, and unintended side effects or extra work.

*2) Labels Are Probabilistic:* As prescribed by the rubric, raters used 1 to encode "definitely included TD" and used 0 to encode "definitely not TD." In a slight modification to the rubric, however, we used any number between 0 and 1 to indicate gradations of certainty. Progressing though the decision tree provided increasing evidence to score tickets as TD items with a confidence measure between 0 and 1. For example, a label of 0.2 indicates that a ticket seemed to contain some mild hints of TD, but most likely is not primarily about TD.

We manually labeled a total of 1,934 tickets. Of these, 441 (32.8%) tickets had a label greater than 0.5. Fig. 1 summarizes the distribution of labels in greater detail.

*3) Inter-rater Reliability:* Our initial batch of 163 tickets was labeled with multiple raters with an inter-rater agreement of 90% (prior to our subsequent reconciliation of disagreements). Subsequent tickets were labeled by one of two experts, where the second expert performed random spot checks. Given our limited budget for producing expert labels, we did not have multiple experts continue to label tickets to produce additional standard inter-rater reliability statistics. We did, however, compute the average prediction error of our main model (which was blind to the rater identity) separately for each rater's collection of tickets and verified that none of the raters displayed a notable tendency to systematically over- or underestimate TD relative to the others.

*D. Deciding Which Tickets to Label (Active Learning)*

We used preliminary versions of the classifier to help decide which Chromium tickets to manually label next. After obtaining new labeled tickets, we retrained the classifier on the now-larger training data and used this new classifier to help recommend yet another set of tickets to label. (This iterative process is known as active learning in the machine learning literature [27].)

Our ticket selection criteria changed over time as our classifier grew more sophisticated. We initially sought examples of TD by performing basic keyword searches on terms like "debt" or "compatible." Later we began to consider how to optimally select new tickets to label on the basis of which labels might most improve the overall performance of successive classifiers trained on the new labels. Our sampling strategy on the final batch was a weighted random sample with (a) weights proportional to the classifier's estimated probability that a ticket contains TD but (b) with a lower bound on the weight to ensure a minimal rate of continued sampling among classes of tickets for which the classifier could have become overconfident that TD does not exist.

Fig. 2 shows the cumulative sum of TD labels (probabilities) over the sequence of labeled tickets, ordered by the time that we created the label. The slope of the curve aligns well with changes in our active learning goals over time. Initially we were trying to get more examples of TD. We got gradually better, illustrated by the increasing slope. But around ticket 800, we became concerned about under-sampling from the population of non-TD tickets. We began to deliberately sample from tickets that our model rated as having a low TD probability, accounting for the decrease in slope.

## III. FEATURE ENGINEERING

Often the most labor-intensive part of any classification problem is to generate and test features that may be relevant to the classification target. We present here a list of features that we tested. In the end, our main model used only a small fraction of these features. Section V discusses which features turned out to be the most useful.

We generated many of our features from free text. We defined the free text of a ticket as the concatenation of its title,

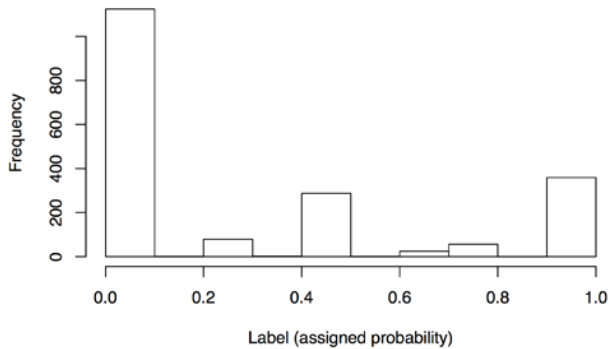

Fig. 1. Distribution of manual labels.

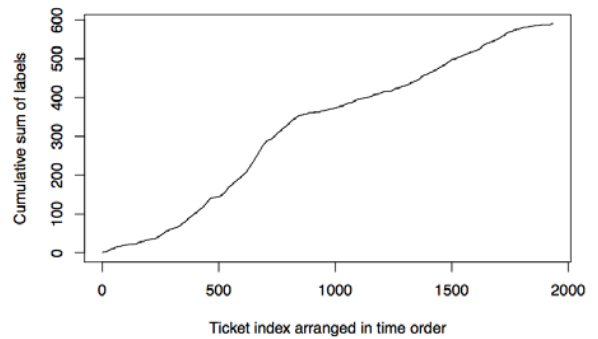

Fig. 2. Sum of ticket labels over order of increasing time.

description, and all comments.

*A. Ticket Metadata*

Each ticket had several fields associated with it. Four fields were used in this study:

- **Authorship:** An authorship field indicates the primary author of each ticket as a sometimes-partially-obfuscated developer email address. From this we defined four authorship flags indicating whether the author email address belongs to one of the following most-common email domains: chromium.org, gmail.com, google.com, and etouch.net. In addition, a "role" feature indicates whether the primary author is a "Project Member."

- **Priority:** Three priority fields indicate whether the ticket priority field is 1, 2, or 3.

- **Status:** Eight status flags indicate whether the status is WontFix, Fixed, Duplicate, Verified, Archived, Assigned, Available, or Untriaged.

- **Type:** Three type flags indicate whether the type field begins with "bug," "bug-," or some string that does not start with "bug."

*B. Counts*

Count characterizes the length of comments associated with a ticket. Several features reflect counts of words, characters, or other items in the free text:

- number of characters
- number of characters in the longest sentence
- the median number of characters per word after removing HTML
- the number of words in the cleaned text (using the cleaning process described for *n*-grams later in the section)
- number of words in the text after removing HTML
- average number of words per sentence in the cleaned text
- average number of words per sentence after removing HTML
- number of sentences
- number of sha1 hashes (typically associated with discussion of git commits)

*C. Key Phrases*

Key phrases are flags for the presence of at least one exact string match for each key phrase in "debt," "hack," "workaround," "cleanup," "clean-up," "clean up," "give up," "problematic," "not up to date," "inconsisten" [sic], "short term," "deviate," "tweak," "mess," "buggy," "complex," "doesn't work," "out of date," "insufficient," "rework," "remove," "redesign," "refactor," "depend," and "structure." We generated this list of key phrases as potentially indicative of TD based on practical experience studying TD and a review of the literature. Note that "inconsisten" is incomplete intentionally so that it matches both "inconsistent" and "inconsistency."

*D. N-grams*

We used the `quanteda` package in the R scripting language to clean and tokenize the text. This involved converting all characters to lowercase; removing numbers, punctuation, separators (such as tabs), symbols (such as the @ sign), URLs, hyphens, and stop words (as per the English words specified in the stop words R package of type "Snowball" and "Smart"); and applying the `wordStem` package in R to reduce every remaining word/token to its stem. The output of this cleaning process was a vector of words for each ticket. We recorded the number of words in this vector as an additional feature. We then exhaustively computed the *n*-grams for $n = 1, 2,$ and 3, generating tens of thousands of new features. We excluded *n*-grams that were extremely rare or ubiquitous (i.e., those that appeared in fewer than three labeled tickets or were absent in fewer than three labeled tickets).

*E. Concept Words*

We defined a collection of "concept word" features as a kind of continuous generalization of a 1-gram to represent how closely the words of a document approach a target concept. For example, the concept word feature for "outdated" is equal to 1 if "outdated" appears in the ticket, just the same as for the "outdated" 1-gram. Suppose, however, that "old" appears in the ticket while "outdated" does not. Then the "outdated" 1-gram feature is 0 while the concept word "outdated" is substantially greater than 0 (since "outdated" has a meaning closely related to "old").

To compute the similarities between words, we applied the cosine similarity on the 100-d numeric word vectors in the pre-trained `GloVe` word embedding (a vector-space representation of words based on a 6-billion word sample of Wikipedia, available at https://nlp.stanford.edu/projects/glove/). We computed this similarity between each distinct word of a ticket against a concept word and recorded the maximum of these similarities as the concept word feature. We computed the concept word feature for each of the following target words that we judged to be potentially related to TD: "deviate," "outdated," "redundant," "redesign," "decouple," "complicated," "regret," "corrupt," "horrible," and "delay."

*F. Word Vectors*

We ran the `gensim` [40] implementation of `word2vec` [37] with Continuous Bag Of Words (CBOW) to train a 10-dimensional numeric vector to represent each of the 5,000 most-common words in the free text. We trained this embedding on all of the Chromium text—not only the labeled tickets. The goal of the word2vec algorithm (essentially a shallow neural network) is to assign word vectors such that two words have a high cosine similarity if and only if they tend to be used in similar ways.

We then formed the word vector matrix for each ticket. If a ticket contains *k* words (after deleting all words that are not among the 5,000 most-common words), then the word vector matrix has *k* rows and 10 columns; the *i*th row is the word vector of the *i*th word. Finally, we computed several features based on the word vector matrix:

- For each column (i.e., word embedding dimension), compute the 5th percentile of the values in the column. This produces 10 features.
- For each column, compute the 95th percentile of the values in the column. This produces 10 features.
- Compute all of the Euclidian distances between successive rows; summarize this set of distances with two features: the 5th percentile and the 95th percentile. These "sequential differences" features can be viewed as representations of the free text's "twistiness" through the word vector space.

*G. Document Vectors*

We ran the `gensim` implementation of `doc2vec` [30] to train a 20-dimensional document embedding, treating each ticket as a document. This directly produced 20 new features. Document vectors are particularly abstract with no simple interpretation. Conceptually, the goal of `doc2vec` is that the embedding vectors for two different tickets should be similar in proportion to the degree of semantic similarity between the two tickets.

IV. AUTOMATICALLY DETECTING TECHNICAL DEBT

We developed a method to automatically detect TD. This broadly consisted of the following tasks:

(1) Decide which tickets to label (Section II).
(2) Engineer features that are relevant to TD (Section III).
(3) Select a classifier and optimize it according to various measures of performance (this section).

The development process was highly iterative and experimental. Several metrics influenced our choices of which classifier, hyperparameters, and features to use. Since such an iterative process can lead to overfitting the model too closely to a particular set of ticket data, we deliberately performed our final evaluation on a holdout test set of labeled tickets to provide a fair assessment of model performance (Section V).

*A. The Main Model*

We used the R language application programming interface for the LightGBM classifier. LightGBM is a Microsoft-hosted open-source implementation of gradient-boosting machines (GBMs) widely regarded as a slightly improved successor to XGBoost, a GBM implementation that won acclaim in numerous data science competitions such as those hosted by kaggle.com [26].

LightGBM works by constructing a sequence of decision trees that estimate the probability of TD in terms of the features that we provide to the algorithm. Each tree individually is a weak classifier, but summing predictions over a large number of trees tends to improve performance. This basic notion of averaging across trees is also a key to the success of the random forests classifiers. Unlike random forests, however, GBMs typically involve significantly fewer trees, achieving greater statistical efficiency by constructing the trees sequentially and training each new tree based on the errors in the existing ensemble.

We constructed our main model by applying LightGBM with the "xentropy" objective function, which uses the log loss metric to do logistic regression with probabilistic labels. We manually tuned the hyperparameters to values that were generally quite close to the defaults, but we tended to favor oversmoothing to help ensure that the model would continue to perform well beyond the training set. Specifically, our non-default hyperparameter settings consisted of using 60 trees, a maximum of 9 leaves in each tree, a minimum of 10 tickets in each leaf, and a learning rate of 0.04. We omit any further details on LightGBM hyperparameters, as package is documentation is readily available [36].

We used cross-validation to test LightGBM with various combinations of the features described in Section III. For our final model, we excluded all *n*-grams due to overfitting concerns (we had about 50,000 *n*-grams) and because they did not appear to significantly improve performance. We kept all of the remaining 108 features, and 62 of these ended up appearing in at least one of the boosted trees.

Fig. 3 shows an example of the types of trees LightGBM builds on our feature set. To interpret each tree, start from the top and work down: given a new ticket, the tree says that the first thing to look at is "KEYPHRASE_debt," a flag for the presence of the string "debt" in the ticket. If "debt" is present, go right; otherwise go left. Looking at the left-side branch, "n_word_clean" is the number of words in a cleaned version of the ticket text. If there are more than 764 words, go right, and so on. Each leaf is a small number. A large (or less negative) number represents a relatively high likelihood that tickets falling into this bucket are TD.

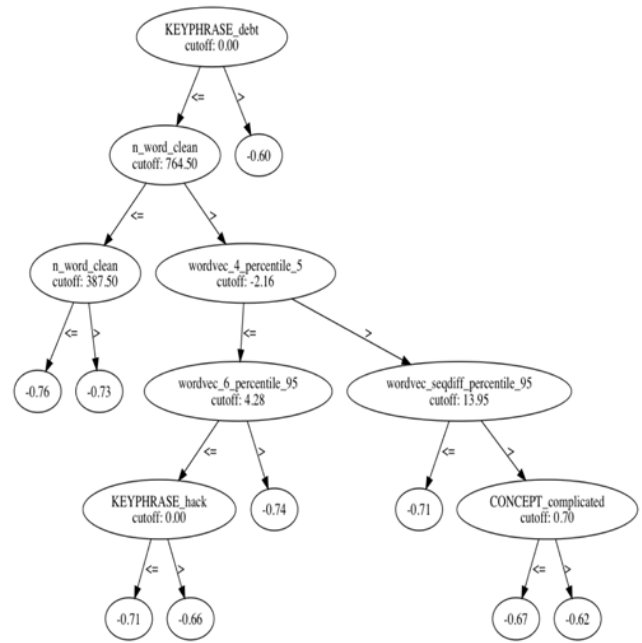

Fig. 3. The first decision tree—out of 60 decision trees in the LightGBM ensemble—comprising our main model.

Applying the sigmoid function (i.e., inverse logistic function) to the leaf-node values produces corresponding estimates of the probability of TD for tickets in those nodes. Specifically, while the leaf-node values in Fig. 3 range from −0.76 to −0.6, the corresponding probability estimates range from 0.319 to 0.354. These estimates are centered so tightly around 0.328 (the average of the training labels) because the low learning rate (0.04) effectively constrains each tree to a relatively narrow range of leaf-node values. However, each tree individually contributes only a small part of the model's overall estimate for each ticket. LightGBM sums a ticket's predicted leaf-node values across all of the trees before applying the sigmoid function to convert this score to an estimated probability that the ticket is TD.

LightGBM does not simply check for the presence of a feature in a ticket. The deep-tree structure of a GBM's ensemble of decision trees can represent complex interactions between numerous factors. Specifically, combinations of features could potentially capture meaning even if individual appearances of those features were irrelevant for TD in isolation.

### B. Baseline Models

We tested two simple baseline models. Our first baseline is a "no TD" classifier, which assigns every ticket the label 0, essentially saying that TD does not exist in all tickets. The main point of considering such an obviously useless classifier is to emphasize the importance of considering not only accuracy but also more detailed metrics such as precision and recall (see Section V-A).

Our second baseline model is a "keyphrase query" classifier that emulates the process that a skilled TD investigator might follow to detect TD at scale. This consists of making a list of key phrases that are suggestive of TD and then assigning a label of 1 for each ticket that contains at least one exact match to any of the key phrases. To implement this, we constructed the list of key phrases in Section III-C in descending order of expected usefulness. We then tuned the keyphrase classifier by querying on only the first $k$ phrases in the list. For example, a large $k$ leads to high recall but poor precision. We chose $k = 12$ to achieve approximately the same ratio of precision to recall as we observed in our main model.

### C. Measuring Classifier Performance

*1) Metrics:* Our performance metrics are accuracy, precision, recall, and AUROC. Accuracy is well known to be a poor metric for classification problems with even slightly imbalanced data; we include it only as a supplemental point of reference. Precision and recall, taken together, offer a more-complete picture than accuracy but fall short in that they force information loss by requiring us to round probabilistic label predictions according to a largely arbitrary probability threshold. Our preferred metric therefore is the AUROC, a nuanced measure of the degree to which tickets with a higher predicted probability of being TD truly are TD. This metric penalizes false positives and false negatives equally and gives greater partial credit for a predicted label of 0.9 than a predicted label of 0.8 when the expert rater label is 1.

*2) Label Uncertainty Weights:* While it is standard for a classifier to output probabilistic labels, it is perhaps less common for the ground truth labels to also be probabilistic (Fig. 1). We defined a label uncertainty weight $w2_i$ for the $i$th expert label $y_i$ to reflect the gradations of certainty in the label. Specifically, since all of our metrics are computable only for binary expert labels, we must round all expert labels to a corresponding binary value $\tilde{y}_i$ in the set {0,1}, but this introduces a problem. Consider what happens if $y_i = 0.5$: we round it up to $\tilde{y}_i = 1$ and then penalize the classifier if the prediction is $p_i = 0.49$, since the rounded version of the prediction is then $\tilde{p}_i = 0$. But this is not reasonable when $y_i = 0.5$; a prediction of 0.49 is exactly as good as a prediction of 0.51. Therefore, we weight the evaluation based on the certainty of the expert label: $w2_i = (0.5 - p_i)^2$. Notice that this weight approaches 0 as $p_i$ approaches 0.5, a state of total uncertainty.

*3) Sampling Weights:* The purpose of computing metrics for our main model is to estimate how well the model would perform on the vast majority of Chromium tickets, including the 475,000 tickets that we scraped but did not yet label. That is, we want a model that will perform well for the vast majority of Chromium tickets, which are neither in our training set nor in our test set. In ideal circumstances, we would estimate the performance based on a simple random sample in which every ticket has an equal probability of selection. Instead, we deliberately biased our sample to over-represent tickets with TD (Section II-D). Such biasing leads to better balance in the distribution of labels but also creates the need to weight the tickets when computing the metrics [10]. Thus, we estimated sampling weights to correct for sampling bias when evaluating performance. Specifically, we built a logistic regression model (again using LightGBM) to estimate the sampling rate for each of the labeled tickets. The training data for this regression included the same features that we used in our main model and included all of the 477,000 tickets. The response variable was the binary indicator of each ticket being included in the evaluation set (e.g., the set of labeled tickets being used to compute metrics). The regression estimate for each ticket is effectively a smoothed local (in terms of the features) estimate of the fraction of similar tickets that are included in the evaluation set. The reciprocal of the sampling rate (minus 1) is approximately the number of closely related tickets that did not get sampled corresponding to each ticket that did get sampled. Thus we defined the $i$th sampling weight $w2_i$ as the reciprocal of the estimated sampling rate.

*4) Combined Importance Weights:* We defined the combined importance weight as the product of the label uncertainty weight and the sampling weight: $w_i = w1_i * w2_i$. Fig. 4 shows the distribution of combined importance weights across the 588 tickets in our holdout test set.

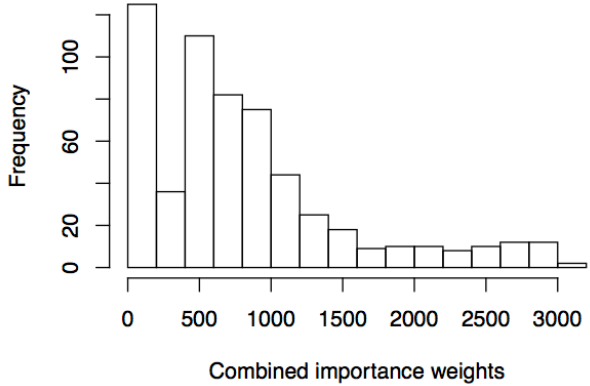

Fig. 4. The distribution of the combined importance weights.

*5) Weighted Metrics:* We defined a weighted version for each of our metrics to incorporate the combined importance weights. For example, we replaced the standard formula for precision with

$$Precision = \frac{\sum_i w_i \tilde{p}_i \tilde{y}_i}{\sum_i w_i \tilde{y}_i} \quad (1)$$

This formula equals the standard definition of precision if all the weights are equal. The weighted generalizations of recall and accuracy are analogous to that for precision. The formula for weighted AUROC is by far most complicated among our metrics; we relied on the `WeightedROC` package in `R` to compute this.

*6) Confidence Intervals:* Simple closed-form expressions for confidence intervals for our weighted metrics are not available. However, we can compute approximate confidence intervals with a nonparametric bootstrap procedure (e.g., see [16]) as follows:

1. Sample the tickets uniformly *with replacement*. This results in a new data set that omits some of the tickets in the original set and includes other tickets multiple times.
2. Compute each metric on the new data.

Repeating this procedure 500 times produces 500 bootstrap replicates of each metric. Finally, the 0.025 and 0.975 quantiles of the 500 replicates provide an approximate 95% confidence interval for each metric.

*7) Cross-validation:* During model development (but not in the final performance evaluation, Section V-A), we computed each of the metrics above using 10-fold cross-validation. Specifically, we randomly split the labeled data into 10 disjoint folds and rebuilt each model 10 different times to predict on each holdout fold.

## V. ANALYSIS AND RESULTS

### A. Classifier Performance

Classifier development concluded when we had a total of 1,346 tickets. We then labeled an additional 588 tickets to serve as a "pure" holdout test set for evaluating classifier performance. The assessment in this section is based entirely on this holdout test set.

TABLE I. UNWEIGHTED PERFORMANCE METRICS

|  | Accuracy | Precision | Recall | AUROC |
|---|---|---|---|---|
| no TD | 0.67 | NA | 0.00 | 0.50 |
| keyphrase query | 0.69 | 0.55 | 0.32 | 0.60 |
| main model | 0.72 | 0.57 | 0.60 | 0.77 |

Table I summarizes the unweighted performance metrics of our main model and the baselines. The accuracy of the "no TD" baseline simply reflects that about $1 - 0.33 = 0.67$ is the fraction of tickets in the holdout test set that had a label less than 0.5. The main model performed modestly better than the keyphrase query baseline on all metrics. As per Section IV-C, however, the unweighted metrics suffer from sampling bias. We include Table I only as an extra reference point.

Table II summarizes the weighted performance metrics of our main model and both baselines. The accuracy of the "no TD" baseline is even higher than in Table I because the non-TD issues were relatively under-represented in our test set, so the average sampling weight for non-TD tickets is considerably greater than the average for TD tickets. Consistently with Table 1, the main model outperformed the keyphrase query baseline.

Table III shows the bootstrapped lower and upper bounds of a 95% confidence interval for the improvement of the main model over the keyphrase query baseline.

Fig. 5 displays the cumulative recall curve (also known as a cost-effectiveness curve) for each model [4]. This curve aligns closely with how one might use the predictions of a model to identify as much TD as possible on a limited budget. First the tickets are sorted in descending order of estimated probability of concerning TD to form a queue. The user then proceeds through the queue, verifying or otherwise processing each ticket. The curve shows the estimated cost-effectiveness of this procedure in terms of the number of TD tickets identified versus the number examined.

TABLE II. WEIGHTED PERFORMANCE METRICS

|  | Accuracy* | Precision* | Recall* | AUROC* |
|---|---|---|---|---|
| no TD | 0.90 | NA | 0.00 | 0.50 |
| keyphrase query | 0.83 | 0.26 | 0.35 | 0.62 |
| main model | 0.87 | 0.40 | 0.62 | 0.88 |

\* We used a weighted version of each metric (see Section III).

TABLE III. CONFIDENCE INTERVAL[†]

|  | Δ Accuracy | Δ Precision | Δ Recall | Δ AUROC |
|---|---|---|---|---|
| 2.5% | −0.00056 | 0.044 | 0.16 | 0.19 |
| 97.5% | 0.07600 | 0.240 | 0.38 | 0.32 |

† Each column displays a 95% confidence interval for the increase in performance between the keyphrase query and the main model.

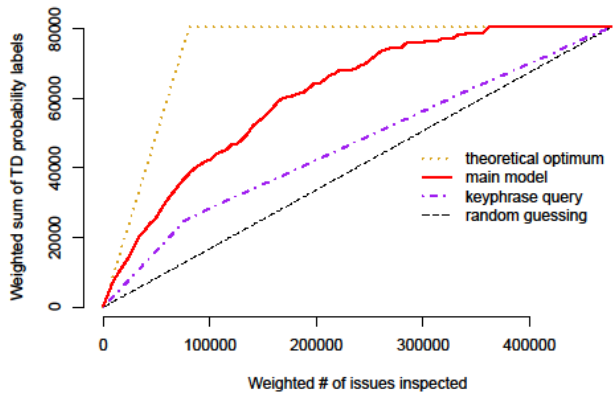

Fig. 5. Cumulative recall plots for the holdout set.

The theoretical optimum cost-effectiveness curve corresponds to perfect sorting, in which all TD tickets appear before any non-TD tickets. With a random sort order—"random guessing"—one would encounter cases of TD at a constant rate on average. Keyphrase query and our main model both do considerably better than random guessing but considerably worse than the theoretical optimum.

*B. Feature Importance*

LightGBM used 62 of the 108 features supplied. Table IV shows the top 15 features based on information gain, summed across all branches of all trees. The precise ordering of these features is not particularly meaningful, as multicollinearity can lead to highly important features contributing little information gain. For example, the key phrase terms "depend" and "redesign" relate to the manual TD-labeling decision point that asks if there is a design issue or limitation underlying a request to fix a defect or implement a new feature, but neither of these key phrases ranks highly. Generally, however, the highly ranked features provide a relatively large fraction of the model's predictive power.

The feature wordvec_4_percentile_5 was an especially strong predictor. As per Section III-F, the number 4 in the feature refers to the 4th dimension of the word2vec embedding. The 5th percentile refers to the lower end of the range of values on that dimension. Sorting the vocabulary of embedded words by their 4th dimension and looking at the bottom 6 words in that list, we find "reapply," "refers," "editissue," "reviewed," "evaluated," and "refs." It should be no surprise that this corner of word space is associated with TD, even though none of these words appeared in our manually generated lists of key phrases and concept words.

The key phrase "debt" unsurprisingly appears among the most important features, and its strong importance suggests that developers who mention "debt" on issue trackers are typically discussing TD and nothing to do with financial loans.

TABLE IV. TOP FEATURES

| Rank | Feature | Information Gain |
|---|---|---|
| 1 | wordvec_4_percentile_5 | 0.262 |
| 2 | KEYPHRASE_debt | 0.214 |
| 3 | n_word_clean | 0.075 |
| 4 | n_sent | 0.075 |
| 5 | wordvec_8_percentile_5 | 0.037 |
| 6 | CONCEPT_deviate | 0.031 |
| 7 | wordvec_6_percentile_95 | 0.017 |
| 8 | wordvec_9_percentile_5 | 0.015 |
| 9 | n_word_no_html | 0.015 |
| 10 | wordvec_3_percentile_5 | 0.014 |
| 11 | docvec_14 | 0.013 |
| 12 | wordvec_1_percentile_5 | 0.012 |
| 13 | avg_nword_clean_per_sent | 0.012 |
| 14 | KEYPHRASE_workaround | 0.011 |
| 15 | docvec_10 | 0.011 |

CONCEPT_deviate is an important concept word feature (Section III-E). A value less than 1 may indicate the presence of similar words such as "differ," "vary," or "change."

Two other important features describe the quantity of free text. The "n_word_clean" feature is the number of words after we cleaned the text (removing stop words and punctuation, etc.). The "n_sent" feature is the number of sentences. They relate to the labeling rubric criterion of whether there is sufficient information to determine that developers are discussing TD. Very short tickets are often marked with the status of "Duplicate" (where the evidence resides in another ticket), "WontFix" (where the submitter has provided insufficient information for the developer to understand the problem), or "Archive" (where the ticket hasn't attracted sufficient interest for the developers to discuss). However, there are exceptional cases where TD is stated concisely.

Table V shows the number of features grouped by feature type and ranked by the total information gain. The first five feature types are represented in the top features (Table IV). Ticket field metadata such as priority and author, however, are not.

TABLE V. FEATURE TYPES RANKED BY IMPORTANCE

| Rank | Feature Type | Number of Features | Total Gain |
|---|---|---|---|
| 1 | word vectors | 19 | 0.407 |
| 2 | key phrases | 9 | 0.262 |
| 3 | counts | 6 | 0.184 |
| 4 | concept words | 10 | 0.072 |
| 5 | document vectors | 15 | 0.070 |
| 6 | priority | 1 | 0.003 |
| 7 | author | 2 | 0.002 |

## C. The Number of Chromium Tickets That Contain Technical Debt

We estimate the number of Chromium tickets, among the 477,000 that we scraped, that concern TD. We start by estimating the fraction $r$ of the tickets that concern TD. A naïve way to do this is to simply compute the rate for the sample of 1,934 tickets:

$$\hat{r}_{\text{naïve}} = \frac{\sum y_i}{1934} \approx 0.3 \quad (2)$$

As discussed in Section IV-C, however, this estimate is subject to sampling bias, so we wish to appropriately weight the sum with sampling weights. Here again we use LightGBM to estimate a logistic regression where the training data includes all of the tickets, and the response variable is the flag indicating whether each ticket is among the 1,934 labeled tickets. We then apply the fitted probability $p_i$ for the $i$th labeled ticket with label $y_i$ to get

$$\hat{r} = \frac{\sum p_i y_i}{\sum p_i} = 0.161 \quad (3)$$

Thus, approximately 16.1% (or about 76,900) of the tickets concern TD. Bootstrapping this estimation procedure over all 477,000 tickets leads to a 95% CI of (14.5, 17.1)% or (69,000, 81,800) tickets.

## VI. THREATS TO VALIDITY

We identify the following potential limitations of our model.

*Manual labeling accuracy:* Our manual labeling is based on a rubric developed by Bellomo et al. [8] that leaves significant room for subjective judgement to individual labelers. While inter-rater reliability analysis can partially address this problem, we were able to directly assess inter-rater agreement for only a small subset of the tickets due to our limited labeling budget.

*Sampling weight estimation*: How we estimate sampling weights affects both our performance evaluation (Section IV-C) and our TD prevalence estimate (Section V-C). The best estimation method is not obvious, and alternative approaches can affect the final evaluation. Crucially, under-smoothing produces sampling weights with excessively high variance while over-smoothing reduced the variance but biases the estimated weights. The standard ways of managing the bias–variance trade-off do not apply directly in this case because estimating the weights is only an intermediate step toward performance evaluation, and there are no "ground truth weights."

*External validity:* It is hard to tell how well our results may generalize outside of the Chromium context. On one hand, our analysis used only Chromium tickets, and many of the most interesting tickets have to do with context that is specific to Chromium. On the other hand, Chromium tickets represent a diverse sample of types of software anomalies and issues across the TD landscape including limitations in code, design, and production infrastructure.

## VII. RELATED WORK

Although TD began as a metaphor [35], it is becoming more common to view it as a tangible software development artifact [41]. Interview and survey studies confirm that developers address TD concretely and try to manage it during software development [17] [19] [32] [42]. Systematic studies present taxonomies that decompose TD into subcategories related to software artifacts such as design, code, and testing [1] [24] and map TD to different development stages such as architectural debt management [2] [31] [43].

A number of studies have looked for relationships between software metrics and TD [18] [25] [34]. Zazworka et al. report a case study where developers identify debt that code analysis tools do not [44]. All of these studies demonstrate examples that would result in a recorded issue similar in nature to the kinds of TD issues discussed by the Chromium developers.

Using developer comments to understand TD has been investigated both empirically and with automated techniques. Bellomo et al. studied four different projects to develop a decision tree to differentiate defects, feature requests, and TD issues [8]. The comments that developers leave in the code have also been studied to locate TD, referred to as self-admitted TD [7] [22]. Bavota et al. [7] conclude that while TD is diffused, it increases over time and survives over 1,000 commits on average. Studying eight open-source projects, Huang et al. [22] conclude that developers refer to TD with a certain vocabulary and discuss it. These findings are consistent with ours, indicating the value of understanding TD and developing practices to locate TD during software development.

A significant body of work focuses on mining data from change request and bug databases to detect where issues have occurred in the past and uses that information for improved definitions, quality analysis, development management, and predictive models. Examples include but are not limited to manual and automated mining of issue trackers for misclassification [20], duplicates [9], and correlations of vulnerabilities and bugs [11]. Issue trackers also serve as a source of historical data to help identify patterns to assist with predicting current or future events, such as risks [12].

Several studies have looked at how issues are reported and resolved in the software engineering life cycle to improve their time to resolution and avoid unintended consequences due to misreporting them [21] [45]. A number of studies look at the quality of reported data and ways to improve it, such as ensuring that missing links between bugs and bug-fix commits are included [6]. These studies suggest that reports that contain key information get addressed sooner. Standards such as ISO/IEC 1044-2009 [23] provide guidance for writing bug reports with enough information so they can be reproduced and fixed. These essential properties are encoded in predefined fields in some issue trackers. However, these fields are not sufficient for describing TD.

Herzig et al. [20] observe that 33.8% of reported issues in bug reports are misclassified when relying only on the classification labels and data sets trained as such. Issues categorized based on the text itself perform significantly better [3]. Consequently, text mining techniques have been increasingly used to identify themes to improve accuracy of resolution. For example, topic modeling has been applied to both improving accuracy of bug classification [38] as well as classifying problem reports [29].

Our study takes a text mining approach. We observe that the most significant TD issues require multiple developer discussions on making design changes, consequently the tradeoff between keeping TD or remediating it. These discussions often get buried in the lengthy unstructured text within issue tickets. Applying machine learning techniques to automatically detect TD can be extremely useful, and is almost a necessity, under such conditions. To the best of our knowledge, this study is the first to do so.

## VIII. CONCLUSIONS AND FUTURE WORK

Machine learning can help (a) detect discussions of TD at scale and (b) identify features that are strongly associated with TD and thus potentially useful for understanding or defining TD. Free text in the Chromium project alone points to many tens of thousands of TD discussions, suggesting that studies of TD have only begun to scratch the surface of a large class of code development challenges.

This paper demonstrates one way to apply GBMs to a sample of labeled tickets from the Chromium project issue tracker. Through an iterative process of improving feature selection and model tuning, we were able to obtain classification AUROC of 88%. Moreover, we estimate that TD may account for about 16% of Chromium tickets. Our future work will refine our main model by engineering better features and increasing the amount of labeled data.

Our experience and findings from this study inform both software engineering practice and research. Our sample data set of 1,934 labeled tickets is a reference point for anyone trying to refine the definition of TD and provides examples for software engineering teams who would like to experiment with more explicitly tracking TD along with other software anomalies such as bugs and vulnerabilities. The tickets identified as TD by our classifier can be studied further and additional features, over and above those used in our model, can be tested by other researchers to develop new automatic TD detection methods.

The extent of TD identified in a large project, such as the one used in our study, suggests augmenting issue trackers with fields that can effectively monitor TD in large projects. Such information can be invaluable to project managers when making decisions on allocating resources to new feature development or paying down TD. If we can effectively classify TD-related comments and issues, we can focus on what practices could be most useful for its timely communication and resolution.

The most ambitious goal of this study was to produce a perfect classifier that automatically determines whether an arbitrary ticket in an issue tracker relates to TD. Such an oracle would

- support any TD management strategy by providing a list of all TD-related tickets in a repository.
- obviate the need for developers to manually apply a TD ticket label and for analysts to be trained in a standard TD terminology (provided they operationally understand it well enough to talk about it in some language).
- practically define TD, to the extent that the oracle is human-interpretable, providing reasoning behind each classification.

While our classifier represents progress, it falls far short of an oracle. We expect future work in two key areas to significantly improve its accuracy, increasing the amount of labeled data (which is not as readily available from public sources such as GitHub as one might expect) and improving feature engineering. Increasing the amount of labeled data via the current post hoc manual process is costly, but other label sources may soon become more common, particularly as more projects begin to use TD as a standard issue tag. Feature engineering advances are needed as well, and the problem is nontrivial. In the current work we tried most of the obvious feature engineering strategies in natural language processing, and fundamental advances may be needed before additional features can yield a sizeable performance boost.

## ACKNOWLEDGMENT

Copyright 2018 IEEE. All Rights Reserved. This material is based upon work funded and supported by the Department of Defense under Contract No. FA8702-15-D-0002 with Carnegie Mellon University for the operation of the Software Engineering Institute, a federally funded research and development center. References herein to any specific commercial product, process, or service by trade name, trade mark, manufacturer, or otherwise, does not necessarily constitute or imply its endorsement, recommendation, or favoring by Carnegie Mellon University or its Software Engineering Institute. DM18-0447.

We thank April Galyardt and Tamara Marshall-Keim for their valuable feedback and expert input.

## REFERENCES

[1] Alves, N. S. R., Ribeiro, L. F., Caires, V., Mendes, T. S., and Spínola, R. O. 2014. "Towards an ontology of terms on technical debt." ACM SIGSOFT, 40, 2, pp. 32–34.

[2] Ampatzoglou, A., Ampatzoglou, A., Chatzigeorgiou, A., and Avgeriou, P. 2015. "The financial aspect of managing technical debt: A systematic literature review." Inform. Software Tech., 64, pp. 52–73.

[3] Antoniol, G., Ayari, K., Di Penta, M., Khomh, F., and Guéhéneuc, Y.-G. 2008. "Is it a bug or an enhancement? A text-based approach to classify change requests." In Proceedings of the 2008 Conference of the Center for Advanced Studies on Collaborative Research: Meeting of Minds (Ontario, Canada). New York: ACM, pp. 23:304–23:318.

[4] Arisholm, E., Briand, L. C., and Johannessen, E. B. 2010. "A systematic and comprehensive investigation of methods to build and

[4] evaluate fault prediction models." Journal of Systems and Software, 83, 1, pp. 2–17.

[5] Avgeriou, P., Kruchten, P., Nord, R., Ozkaya, I., and Seaman, C. 2016. "Reducing friction in software development." IEEE Software, 33, 1, pp. 66–73.

[6] Bachmann, A., Bird, C., Rahman, F., Devanbu, P., and Bernstein, A. 2010. "The missing links: bugs and bug-fix commits." In Proceedings of the 18th ACM SIGSOFT International Symposium on Foundations of Software Engineering (Santa Fe, NM). New York: ACM, pp. 97–106.

[7] Bavota, G. and Russo, B. 2016. "A large-scale empirical study on self-admitted technical debt." In Proceedings of the 13th International Conference on Mining Software Repositories (Austin, TX). New York: ACM, pp. 315–326.

[8] Bellomo, S., Nord, R., Ozkaya, I., and Popeck, M. 2016. "Got technical debt? Surfacing elusive technical debt in issue trackers." In Proceedings of the 13th International Conference on Mining Software Repositories (Austin, TX). New York: ACM, pp. 327–338.

[9] Bettenburg, N., Premraj, R., Zimmermann, T., and Kim, S. 2008. "Duplicate bug reports considered harmful … really?" In IEEE International Conference on Software Maintenance (Beijing, China). Piscataway, NJ: IEEE Press, pp. 337–345.

[10] Beygelzimer, A., Dasgupta, S., and Langford, J. 2009. "Importance weighted active learning." In Proceedings of the 26th Annual International Conference on Machine Learning (Montreal, Quebec, Canada). New York: ACM, pp. 49–56.

[11] Camilo, F., Meneely, A., and Nagappan, M. 2015. "Do bugs foreshadow vulnerabilities? A study of the Chromium Project." In IEEE/ACM 12th Working Conference on Mining Software Repositories (Florence, Italy). Piscataway, NJ: IEEE Press, pp. 269–279.

[12] Choetkiertikul, M., Dam, H. K., Tran, T., and Ghose, A. 2015. "Characterization and prediction of issue-related risks in software projects." In Proceedings of the 12th Working Conference on Mining Software Repositories (Florence, Italy). Piscataway, NJ: IEEE Press, pp. 280–291.

[13] Chromium Issue examples to be released with publication.

[14] Chromium Issues. https://code.google.com/p/chromium/issues/list

[15] Codabux, Z. and Williams, B. J. 2016. "Technical debt prioritization using predictive analytics." In Proceedings of the IEEE/ACM 38th International Conference on Software Engineering Companion (Austin, TX). New York: ACM, pp. 704–706.

[16] Efron, B. and Tibshirani, R. J. 1993. An Introduction to the Bootstrap. New York: Chapman & Hall/CRC.

[17] Ernst, N., Bellomo, S., Ozkaya, I., Nord, R. L., and Gorton, I. 2015. "Measure it? Manage it? Ignore it? Software practitioners and technical debt." In Proceedings of the 10th Joint Meeting of the European Software Engineering Conference and the ACM SIGSOFT Symposium on the Foundations of Software Engineering (Bergamo, Italy). New York: ACM, pp. 50–60.

[18] Fontana, F., Ferme, V., and Spinelli, S. 2012. "Investigating the impact of code smells debt on quality code evaluation." In Third International Workshop on Managing Technical Debt (Zurich, Switzerland). Piscataway, NJ: IEEE Press, pp. 15–22.

[19] Guo, Y., Seaman, C., Gomes, R., Cavalcanti, A., Tonin, G., DaSilva, F., et al. 2011. "Tracking technical debt: An exploratory case study." In Proceedings of the 27th International Conference on Software Maintenance (Williamsburg, VA). Piscataway, NJ: IEEE Press, pp. 528–531.

[20] Herzig, K., Just, S., and Zeller, A. 2013. "It's not a bug, it's a feature: How misclassification impacts bug prediction." In Proceedings of the 2013 International Conference on Software Engineering (San Francisco). Piscataway, NJ: IEEE Press, pp. 392–401.

[21] Hooimeijer, P. and Weimer, W. 2007. "Modeling bug report quality." In Proceedings of the 22nd IEEE/ACM International Conference on Automated Software Engineering (Atlanta, GA). New York: ACM, pp. 34–43.

[22] Huang, Q., Shihab, E., Xia, X., Lo, D., and Li, S. 2018. "Identifying self-admitted technical debt in open source projects using text mining." Empirical Software Engineering, 23, 1, pp. 418–451.

[23] IEEE Std 1044-2009. 2009. IEEE Standard Categorization for Software Anomalies. Washington, DC: IEEE Computer Society.

[24] Izurieta, C., Vetro, A., Zazworka, N., Cai, Y., Seaman, C., and Shull, F. 2012. "Organizing the technical debt landscape." In Proceedings of the Third International Workshop on Managing Technical Debt (Zurich, Switzerland). Piscataway, NJ: IEEE Press, pp. 23–26.

[25] Kazman, R., Cai, Y., Mo, R., Feng, Q., Xiao, L., Haziyev, S., et al. 2015. "A case study in locating the architectural roots of technical debt." In Proceedings of the 37th IEEE International Conference on Software Engineering (Florence, Italy). Piscataway, NJ: IEEE Press, pp. 179–188.

[26] Ke, G., Meng, Q., Finley, T., Wang, T., Chen, W., Ma, W., et al. 2017. "LightGBM: A highly efficient gradient boosting decision tree." In Proceedings of the 31st Conference on Neural Information Processing Systems (Long Beach, CA). Curran Associates: https://papers.nips.cc/paper/6907-lightgbm-a-highly-efficient-gradient-boosting-decision-tree

[27] Krishnakumar, A. 2007. Active learning literature survey. Technical reports, University of California, Santa Cruz. 42.

[28] Kruchten, P., Nord, R. L., and Ozkaya, I. 2012. "Technical debt: From metaphor to theory and practice." IEEE Softw. Spec. Issue Tech. Debt, 29, 6, pp. 18–21.

[29] Layman, L., Nikora, A. P., Meek, J., and Menzies, T. 2016. "Topic modeling of NASA space system problem reports: Research in practice." In Proceedings of the 13th International Conference on Mining Software Repositories (Austin, TX). New York: ACM, pp. 303–314.

[30] Le, Q. V. and Mikolov, T. 2014. "Distributed representations of sentences and documents." ArXiv:1405.4053 [Cs]. In Proceedings of the 31st International Conference on Machine Learning (Beijing, China). JMLR.org: http://arxiv.org/abs/1405.4053

[31] Li, Z., Liang, P., and Avgeriou, P. 2014. "Architectural debt management." In Economics-Driven Software Architecture, I. Mistrik, R. Bahsoon, Y. Zhang, K. Sullivan, and R. Kazman, Eds. San Diego, CA: Elsevier, pp. 183–204.

[32] Lim, E., Taksande, N., and Seaman, C. 2012. "A balancing act: What software practitioners have to say about technical debt." IEEE Software, 29, 6, pp. 22–27.

[33] Maldonado, E., Shihab, E., and Tsantalis, N. 2017. "Using natural language processing to automatically detect self-admitted technical debt." IEEE Transactions on Software Engineering, 43, 11, pp. 1044–1062.

[34] Marinescu, R. 2012. "Assessing technical debt by identifying design AWS in software systems." IBM Journal of Research and Development, 56, 5, pp. 9:1–9:13.

[35] McConnell, S. 2007. "Technical debt." Construx: http://www.construx.com/10x_Software_Development/Technical_Debt/

[36] Microsoft Corporation. 2017. Welcome to LightGBM's Documentation! Retrieved from https://lightgbm.readthedocs.io/en/latest/

[37] Mikolov, T., Chen, K., Corrado, G., and Dean, J. 2013. "Efficient estimation of word representations in vector space." In Proceedings of the International Conference on Learning Representations (Scottsdale, Arizona). CoRR: abs/1301.3781. Retrieved from http://arxiv.org/abs/1301.3781

[38] Park, J., Lee, M.-W., Kim, J., Hwang, S., and Kim, S. 2011. "CosTriage: A cost-aware triage algorithm for bug reporting systems." In Proceedings of the Twenty-Fifth AAAI Conference Artificial Intelligence (San Francisco, CA). Menlo Park, CA: AAAI, pp. 139–144.

[39] Potdar, A. and Shihab, E. 2014. "An exploratory study on self-admitted technical debt." In Proceedings of the 2014 IEEE International Conference on Software Maintenance and Evolution (Victoria, BC, Canada). Piscataway, NJ: IEEE Press, pp. 91–100.

[40] Rehurek, R. and Sojka, P. 2010. "Software framework for topic modelling with large corpora." In Proceedings of the LREC 2010

Workshop on New Challenges for NLP Frameworks (Valletta, Malta). Luxemburg: ELRA, pp. 5–50.

[41] Shull, F., Falessi, D., Seaman, C., Diep, M., and Layman, L. 2013. "Technical debt: Showing the way for better transfer of empirical results." In Perspectives on the Future of Software Engineering. Berlin, Germany: Springer, pp. 179–190.

[42] Spínola, R. O., Zazworka, N., Vetro, A., Seaman, C., and Shull, F. 2012. "Investigating technical debt folklore: Shedding some light on technical debt opinion." In Proceedings of the Third International Workshop on Managing Technical Debt (Zurich, Switzerland). Piscataway, NJ: IEEE Press, pp. 1–7.

[43] Tom, E., Aurum, A., and Vidgen, R. T. 2013. "An exploration of technical debt." J. Syst. Softw., 86, 6, pp. 1498–1516.

[44] Zazworka, N., Spínola, R., Vetro', A., Shull F., and Seaman C. 2013. "A case study on effectively identifying technical debt." In Proceedings of the 17th International Conference on Evaluation and Assessment in Software Engineering (Porto de Galinhas, Brazil). New York: ACM, pp. 42–47.

[45] Zimmermann, T., Premraj, R., Bettenburg, N., Just, S., Schröter, A., and Weiss, C. 2010. "What makes a good bug report?" IEEE Trans. Software Eng., 36, 5, pp. 618–643.